\documentclass{aims}
\usepackage{amsmath}
  \usepackage{paralist}
  \usepackage{graphics} 
  \usepackage{epsfig} 
\usepackage{graphicx}
\usepackage{epstopdf}
 \usepackage[colorlinks=true]{hyperref}
\hypersetup{urlcolor=blue, citecolor=red}
\usepackage{booktabs}
\usepackage{float}

\def\currentvolume{X}
 \def\currentissue{X}
  \def\currentyear{200X}
   \def\currentmonth{XX}
    \def\ppages{X--XX}

\theoremstyle{definition}

\title[Facilitating Team-Based Data Science] 
      {Facilitating Team-Based Data Science: Lessons Learned from the DSC-WAV Project}

\author[Legacy, Zieffler, Baumer, Barr, and Horton]{}


 \email{legac006@umn.edu}
 \email{zief0002@umn.edu}
 \email{bbaumer@smith.edu}
 \email{vbarr@mtholyoke.edu}
 \email{nhorton@amherst.edu}

\thanks{$^*$ Corresponding author: Chelsey Legacy}

\begin{document}
\maketitle

\centerline{\scshape Chelsey Legacy$^*$}
\medskip
{\footnotesize
 \centerline{Department of Educational Psychology}
   \centerline{University of Minnesota}
   \centerline{Minneapolis, MN 55455, USA}
} 

\medskip

\centerline{\scshape Andrew Zieffler}
\medskip
{\footnotesize
 \centerline{Department of Educational Psychology}
   \centerline{University of Minnesota}
   \centerline{Minneapolis, MN 55455, USA}
} 

\medskip

\centerline{\scshape Benjamin S. Baumer}
\medskip
{\footnotesize
 \centerline{Program in Statistical \& Data Sciences}
   \centerline{Smith College}
   \centerline{Northampton, MA 01063, USA}
}

\medskip

\centerline{\scshape Valerie Barr}
\medskip
{\footnotesize
 \centerline{Department of Computer Science}
   \centerline{Mount Holyoke College}
   \centerline{South Hadley, MA 01075, USA}
}

\medskip

\centerline{\scshape Nicholas J. Horton}
\medskip
{\footnotesize
 \centerline{Department of Mathematics and Statistics}
   \centerline{Amherst College}
   \centerline{Amherst, MA 01002, USA}
}

\bigskip

 \centerline{(Communicated by the associate editor name)}

\begin{abstract}
While coursework provides undergraduate data science students with some relevant analytic skills, many are not given the rich experiences with data and computing they need to be successful in the workplace. Additionally, students often have limited exposure to team-based data science and the principles and tools of collaboration that are encountered outside of school.

In this paper, we describe the DSC-WAV program, an NSF-funded data science workforce development project in which teams of undergraduate sophomores and juniors work with a local non-profit organization on a data-focused problem. To help students develop a sense of agency and improve confidence in their technical and non-technical data science skills, the project promoted a team-based approach to data science, adopting several processes and tools intended to facilitate this collaboration.

Evidence from the project evaluation, including participant survey and interview data, is presented to document the degree to which the project was successful in engaging students in team-based data science, and how the project changed the students' perceptions of their technical and non-technical skills. We also examine opportunities for improvement and offer insight to other data science educators who may want to implement a similar team-based approach to data science projects at their own institutions.

\end{abstract}


\section{Introduction}

Data science continues to grow within both academia \cite{Baumer:2015, Cetinkaya-Rundel:2021, Donoghue:2021, Hardin:2015, McNamara:2017} and industry \cite{Bureau-of-Labor-Statistics:2021, KnowledgeWharton:2019}. In industry, where the kind of multifaceted problems that companies face require input from groups of people with diverse sets of skills, experiences, and backgrounds, team-based data science is the norm \cite{Maier:2019, Zhang:2020}. Moreover, the data science staff of many companies now work largely or exclusively remotely---a shift that began before the pandemic but accelerated greatly during it \cite{Limoncelli:2020, Sako:2021}. Over time, industry has evolved practices that facilitate the kind of collaborative programming work necessary for software development. Many of these successful industrial practices are now being ported to data science work, for which collaborative programming is necessary but not sufficient. Academia is playing catch-up, and it remains challenging to replicate within brick-and-mortar institutions the dynamic, team-based work environment that undergraduates will see in their immediate futures.

While the courses and programs in academia prepare students by teaching potentially relevant data analytic skills, it is unclear whether coursework alone is enough to provide students with the experiences with data and computing they need to be successful in tomorrow’s workplace. For example, many data science students lack facility with version control systems, or working with non-traditional data (e.g., images, text) \cite{NASEM:2018}. Moreover, because of the time and grading constraints of the academic environment, students do not often have the opportunity to work on open-ended, substantive problems that require a diverse team of experts and weeks, months, or even years to solve.

In this paper, we describe the Data Science Corps: Wrangle-Analyze-Visualize (DSC-WAV) program, a data science workforce development project funded by the NSF as part of the Harnessing the Data Revolution (HDR) initiative. In the DSC-WAV program, teams of undergraduate students work with a local non-profit organization on a data-focused problem. To facilitate a team-based approach to data science, the project adopted an Agile framework, code review, and collaborative management and coding tools (e.g., Git, GitHub, Trello). We present evidence from the project evaluation, including data from surveys and interviews with students and faculty, documenting the projects' successes in utilizing this approach, as well as opportunities for improvement. In particular, we examine how engaging in this team-based data science project helped develop the participants' understanding of their technical and non-technical skills (e.g., interacting with clients), as well as their sense of agency and empowerment for ``doing'' data science.

\subsection{The DSC-WAV Program}

DSC-WAV is a multi-institution, NSF-funded data science workforce development project that provides teams of undergraduate students the opportunity to partner with local governmental and not-for-profit organizations to work on a data-centric problem (see \href{https://dsc-wav.github.io/www/}{https://dsc-wav.github.io/www/}). This project (now working with its fourth cohort of students) arose from the concern that undergraduate data science students’ course-based data experiences primarily have clearly formulated problems and make use of pre-cleaned and overly simplistic data sets \cite{Nolan:2010}. Moreover, many of these experiences are solo endeavors, and do not prepare students for the team-based nature of data science in the workplace \cite{Condon:2021}. We also note that  many small organizations, especially not-for-profits, have collected data for use in decision making, but lack personnel with the technical training to make sense of the information in these data. In our experiences, the data science tasks these organizations require are primarily data wrangling and visualization, tasks ideal for students pursuing a degree in data science or a related field.

In each cohort, teams of between two and five students work with a community organization on a data-focused problem. The community organization provides the domain knowledge for the project and typically assign a liaison to interact with the student team. One of the students on the team is selected for their prior experience and interest in community engagement (community engagement scholar), and can also sometimes provide domain knowledge. The faculty member who lined up the project also acts as supervisor for the team.

Student participants were recruited via email and in-class announcements. We chose to recruit sophomores and juniors to offer students a data science experience earlier in their academic career. We also felt that first-year students were unlikely to have sufficient training in data science. Among those who applied, selection valued experience (e.g., students with more programming experience), overall academic performance, and commitment to community engagement. We also made efforts to ensure that each team was diverse in a variety of respects. The acceptance rate was between 35\% and 45\% for all four cohorts. 74\% of the participants identified as female or non-binary. Nearly all of the participants, in all cohorts, were sophomores or juniors.

Funding from the grant was used primarily to compensate students (26\% of total budget, via participant stipends) and faculty mentors and staff coordinator (28\% of total budget). Nineteen percent of the costs was devoted to food, travel, and event expenses with the remainder assessment and indirect costs. No compensation was provided to the client organizations.

\section{Team-Based Approach to Data Science}

The prevalence of teams in the workplace has inspired a fair amount of research, primarily related to what makes a team successful or efficient. This work has spanned factors such as team size \cite{Amason:1997, Stewart:2006}, composition \cite{Mannix:2005}, leadership \cite{Burke:2006}, and autonomy \cite{Leach:2005, Mierlo:2006}.

Being able to work in teams is an especially important skill for those pursuing a career in data science \cite{NASEM:2018}. Outside of college or university courses, the complexity of data science projects and shared interests from a diverse set of stakeholders (e.g., clients, marketers, systems engineers, web developers, analysts) make teamwork a daily reality for many data scientists \cite{Kim:2016}. Giving students a meaningful team-based data science experience also means introducing them to the structures and processes that are used to make these types of collaborations successful in the workplace. In the next several sections we describe and review some of the literature related to the structures and frameworks the DSC-WAV program uses to engage students in team-based data science and the team-based skills that the program emphasizes.

\subsection{Agile}

Agile is a flexible, team-based approach to project management and development that has been widely adopted in software development projects \cite{Conforto:2014}. The Agile approach also has strong connections to many educational principles, including constructivism, active learning, and collaboration \cite{Lopez-Alcarria:2019}. Furthermore, its use has been linked to the development of professional competencies such as critical thinking, time management, coping with uncertainty, organization, and self-reflection \cite{Kastl:2018, Razmov:2006, Valentin:2015}. Aside from its educational benefits, Agile has also been quite successful in practice. For example, the Standish Group surveyed thousands of IT projects across many sectors and found that those using Agile methodologies had a success rate that nearly doubled those using a traditional waterfall approach\footnote{See \href{https://en.wikipedia.org/wiki/Waterfall_model}{https://en.wikipedia.org/wiki/Waterfall\_model}} \cite{The-Standish-Group:2018}.

\begin{figure}[htp]
\begin{center}
  \includegraphics[width=\linewidth]{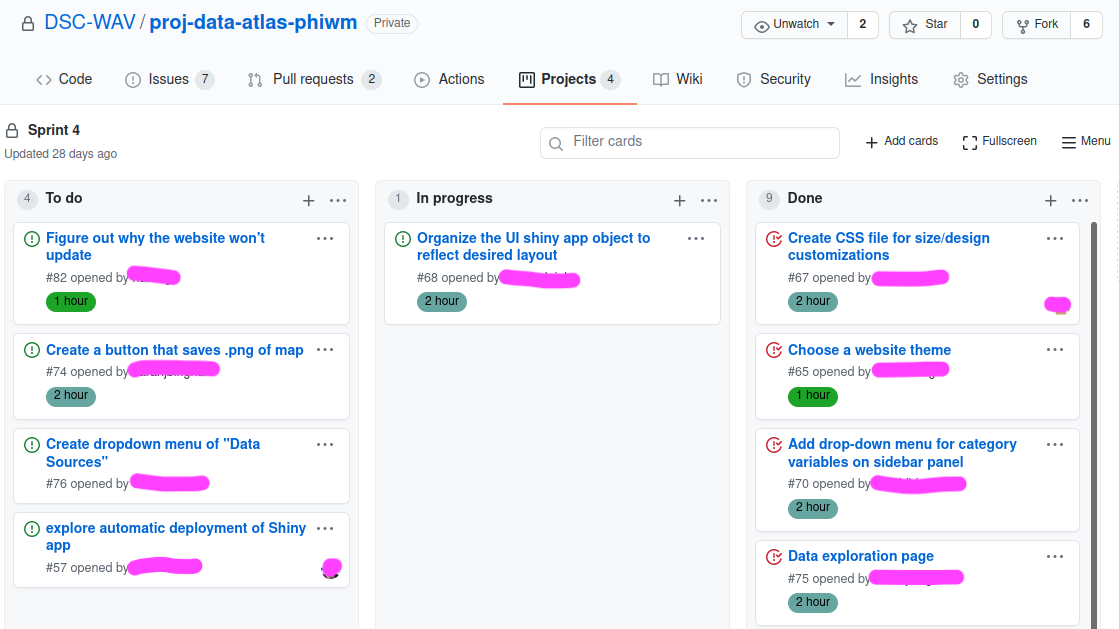}\\
  \caption{A sample Kanban board used by a DSC-WAV team. This board is implemented through GitHub Projects. Each task is represented by a ``card'' and linked to an issue in the GitHub repository. As students work on tasks, they move them from ``To do'' (i.e., the sprint backlog), to ``In progress,'' to ``Done.'' A quick look at the board helps everyone on the team understand who is working on what. This team used tags to estimate the length of time that each task will take to complete.}
  \label{fig:01}
  \end{center}
\end{figure}

There are several methods of Agile implementation \cite{Hossain:2009}. The DSC-WAV program adopted an Agile framework that incorporated elements from two of the most popular methodologies: Scrum and Kanban. From Scrum, we implemented the iterative sprint-based approach to carrying out the project. The project is divided into smaller time-boxed increments called \textit{sprints} (typically lasting 2--4 weeks). Each sprint begins with a \textit{kick-off} meeting in which team members identify detailed goals for the sprint. These are detailed in a \textit{sprint backlog} as \textit{user stories} (which can be subsequently broken down into tasks or issues). Daily meetings, called \textit{stand-ups} ($\sim$15 min), help identify when a user story is completed as well as obstacles preventing completion. At the end of a sprint, the team presents their work to the client in a "review" meeting for feedback. The team also has a \textit{retrospective} in which the work and process are internally evaluated to make improvements in the subsequent sprint. From the Kanban methodology, we adopted the use of a \textit{Kanban board}. Each of the projects' tasks/user stories are represented visually on a digital board, and categorized as “to-do”, “doing”, or “done” (see Figure \ref{fig:01}). As the work on a project progresses, tasks are moved to different categories to update their status.

Several principles and practices for successful Agile implementations have been suggested. These include, for example, a common team vision, development through iteration, and the use of visual artifacts (e.g., boards, panels, and sticky-notes). However, the empirical evidence for whether these practices are correlated with successful project outcomes is less than clear \cite{Qumer:2008, Sheffield:2013, Vazquez-Bustelo:2007}. Some practices, on the other hand, do seem to have more evidential support \cite{Chow:2008, Hoda:2011, Hossain:2009}. For example, there is a fair amount of evidence that teams need to have some degree of synchronous communication (for many Agile projects, team members are co-located). Hossain et al. \cite{Hossain:2009} identify communication issues as the major challenge teams face when adopting the Scrum methodology. In particular, overcoming the lack of synchronous communication is considered a vital obstacle to overcome in distributed teams. Active involvement of the stakeholder throughout the cycle of development is another practice that is empirically linked to project success.

\subsection{Skills Development for Collaborative Data Science}

In addition to the data analytic skills, there are myriad other technical and non-technical competencies that students need to develop as they pursue potential careers in data science. On the technical side, two practices that are assets to students are facility with a version control system (e.g., Git, GitHub) and experience participating in code review (both as reviewer and reviewee). Non-technical skills include collaboration and communication. Both of these have an intra-team component (i.e., collaborating and communicating with other data scientists) and a client component (i.e., collaborating and communicating with subject matter experts who may not be facile with data science).

\subsubsection{Version Control}

Facility with a version control system (VCS) is a crucial technical skill for people engaged in data analysis \cite{Beckman:2021, Nolan:2010, Zolkifli:2018}. The most commonly used VCS in software development is Git, most often used in conjunction with the code hosting site, GitHub \cite{Eghbal:2020}. In addition to Git's usefulness as a tool in the data science workflow---Bryan \cite{Bryan:2018} even suggests its use is `best practice'---the use of Git may promote increased understanding about project management \cite{Hsing:2019}.

\subsubsection{Peer Code Review}

Peer code review is an important aspect of collaboratively writing and maintaining code. Aside from preventing bugs and ensuring that the code adheres to any project guidelines about style and architecture, peer code review also helps coders improve their technical skills by seeing other people's code and solutions \cite{Sadowski:2018}. Peer code review also develops students' ability to read code, a valuable professional skill that facilitates program comprehension \cite{Busjahn:2011, Trytten:2005}. Moreover, it may help students see that code development is not a solitary endeavor in most work environments \cite{Trytten:2005}.

\subsubsection{Collaboration}

Luca and Tarricone \cite[p.~370]{Luca:2001} note that ``there is more to effective teamwork than a keen intellect and grasp of technical knowledge. The difference between success and mediocrity in working relationships, especially in a team environment, can be attributed to a team member's mastery of the [non-technical] skills.'' One essential non-technical skill for team-based data scientists is the ability to collaborate with others toward a shared goal. Processes that facilitate this collaboration include the ability to distribute the resources and workload, negotiate authority, and manage intragroup conflicts that arise \cite{Fredrick:2008}. The research literature suggests that successful teams have some degree of interdependence among members (individual's outcomes are affected by the actions of the other team members) and that teams that build a culture of encouragement and facilitation of team member's efforts perform best \cite{Fisher:1997, Johnson:1995, Johnson:1999, Scarnati:2001}. Other factors also contribute to successful collaboration including self-awareness, empathy, social awareness, motivation, and the ability to inspire and motivate other team members \cite{Goleman:1998, Luca:2001, Yost:2000}.

\subsubsection{Communication}

Successful collaboration requires that team members are able to communicate about the project tasks and goals. Additionally, interpersonal communication among members supports and facilitates team-based work \cite{Harris:1996}. Effective written and oral communication of data-analytic results to both technical and non-technical audiences has also been identified as a component of data acumen necessary for data science students \cite{NASEM:2018, Rawlings-Goss:2018}.

Communication is especially important to consider for distributed teams adopting the Agile methodology, where the ability to communicate synchronously is vital. One potential solution for distributed teams is an instant messaging (IM) communication tool, such as Slack or Microsoft Teams. These communication tools have been shown to improve team cohesiveness, overall communication, and also promote better project planning through the stored history of the team's communication \cite{Zahedia:2016, Zhang:2010}. In addition, IM systems seem to promote ad-hoc conversations which also helps the team to clarify misunderstandings and work more collaboratively to solve project problems \cite{Forsgren:2018}.

\subsection{Team-Based Data Science Implementation in the DSC-WAV Program}

To help engage students in team-based data science over the course of a semester we adopted an Agile framework that incorporated elements from both the Scrum and Kanban methodologies. This included a sprint-based iterative approach in which:

\begin{itemize}
\item The work is organized into a series of short sprints to break up large tasks and help ensure continuous progress and the ability to adapt or modify the project as needed;
\item Subtasks are organized into a backlog to identify priorities for each stage of the analysis. Each subtask is also written as a user story to help students articulate why a particular subtask was needed and it would fit together with other subtasks;
\item The team and faculty advisor meet often (i.e., stand-ups) to update each other on progress and identify roadblocks;
\item At the end of each sprint, results were presented and discussed with the stakeholder (i.e., sprint demo) in the context of the broader goals of the project with adjustments made in advance of the next sprint;  and
\item The team also engages in a sprint retrospective to identify process issues and ways that they might improve their work.
\end{itemize}

\begin{figure}[htp]
\begin{center}
  \includegraphics[width=\linewidth]{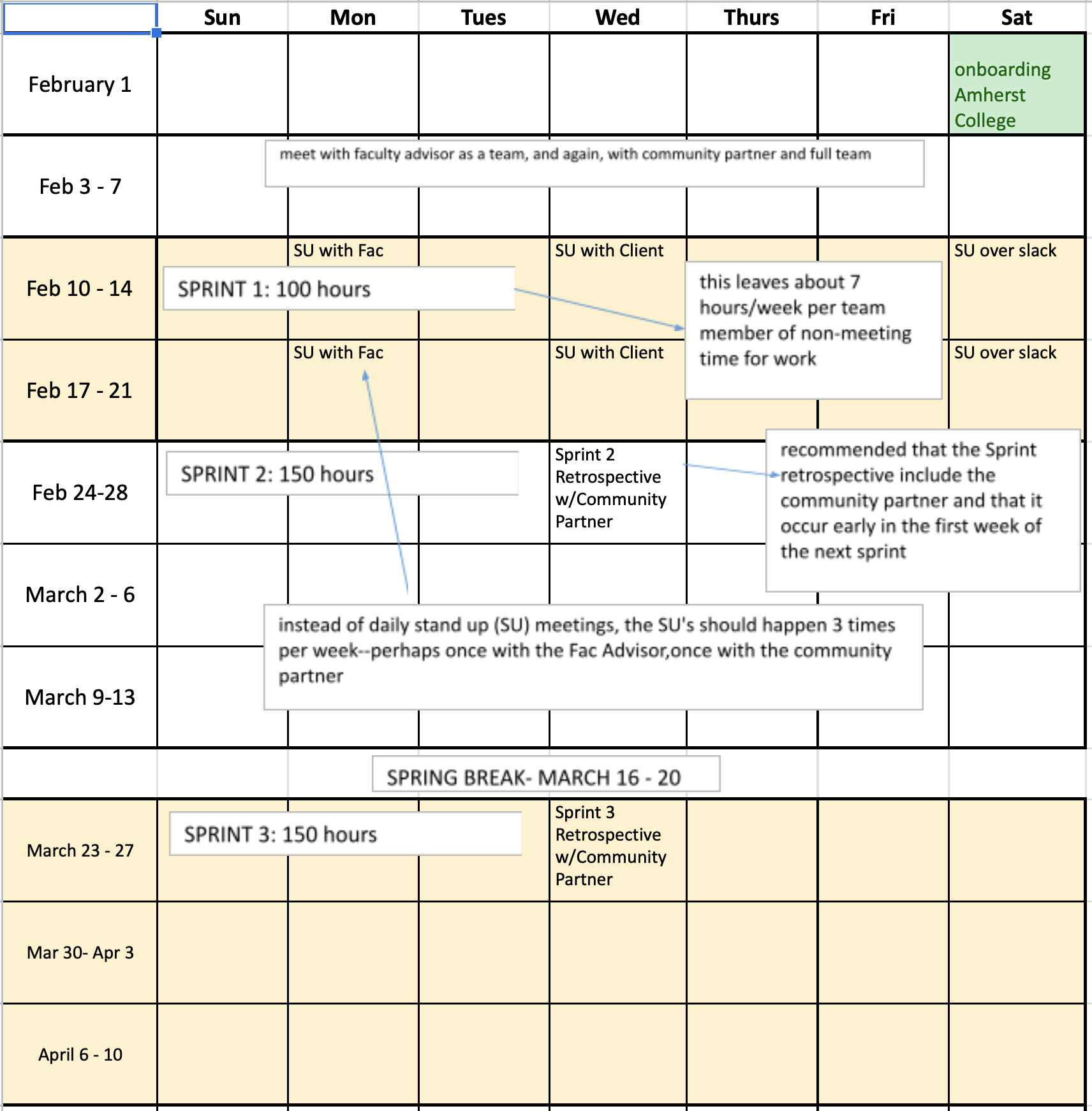}\\
  \caption{Sample timeline for the first six weeks of Cohort 1 that was distributed to teams. A similar schedule was used for later cohorts.}
  \label{fig:02}
  \end{center}
\end{figure}

To promote this iterative approach, the student teams were encouraged to communicate frequently with their community partner and faculty advisor. Figure \ref{fig:02} displays a sample timeline of the first two sprints given to students on a team in Cohort 1 to further reinforce this frequent communication (a similar schedule was used by later cohorts).

The adopted Agile framework also included the use of digital Kanban boards to help teams identify the status of tasks and review team progress and also as a mechanism to promote project management and within-team communication. Additionally, we advocated the use of several other tools and processes to help facilitate and improve students’ administration of the project and improve their professional skills.

\begin{itemize}
\item Slack to foster open team communication\footnote{To promote inclusion and improve communication, students were encouraged to keep all project-centered communication on Slack or GitHub where all group members could see it, and avoid the use of private messaging.};
\item Git and GitHub as a VCS to collaborate on and maintain code and tasks for the project;
\item Code review to help students improve their coding skills and to ensure that the project’s codebase is coherent and works in synchrony;
\item Designation of a ``community-engagement scholar,'' whose primary responsibility is liaising with the project sponsor; and
\item Creation of and sharing a mid-semester update (presentation) and final report of the team's work with the community partner and the DSC-WAV program leadership.
\item Templating in user stories (``As a \_\_\_, I want \_\_\_, so that I can \_\_\_'') and sprint retrospectives (``I like/I wish/What if?'') to develop communication skills.
\end{itemize}

We felt that implementing this set of tools and processes would not only improve the chances of successful collaboration between all stakeholders, but also help students develop a sense of agency and improve their confidence in their technical and non-technical skills for doing data science.

To introduce students to these processes and tools, each cohort of student participants participated in an onboarding workshop prior to beginning the project. The onboarding workshops varied in format and duration for each cohort, but always included an introduction to the DSC-WAV program and content related to the adopted Agile framework, use of GitHub, and client communication. Each workshop also included student-centered activities designed to engage the students, establish team chemistry, and build a shared vision of each team’s project. For additional information about these workshops, see Appendix A.

\subsection{Research Questions}

To study the potential of this type of organizing framework for team-based data science projects, we examined program evaluation data to answer two research questions:

\begin{enumerate}
\item How do undergraduate project teams actually implement Agile as they engage in a team-based data science project?
\item To what extent did the DSC-WAV program's organizing framework support undergraduate students in developing their technical and non-technical data science skills?
\end{enumerate}

\section{Methods}

The data to answer the research questions comes from student responses to end-of-project surveys and interviews. All interviews and surveys were conducted by the first and second author (and a graduate student) who were external evaluators for the DSC-WAV program. The survey items and interview protocols were written using item-writing guidelines \cite{Dillman:2014}, and were intended to collect information about the experiences and perceptions of the stakeholders as they engaged in the project. They also included questions aimed to elicit students' reflections on their experiences working on the project, as well as their perceptions about their own personal growth as a data scientist and collaborator. All instruments are available at \href{https://dsc-wav.github.io/www/outreach.html}{https://dsc-wav.github.io/www/outreach.html}.

Interviews were conducted with almost all of the student participants after the first cohort and with a subset of student participants from the third cohorts. These interviews, which lasted approximately 30--45 minutes, were conducted via Zoom and were audio recorded. No interviews were conducted at the end of Cohort 2 (Fall 2020) due to complications related to the pandemic. The end-of-project survey was administered via Qualtrics to all participants in Cohort 2 and to those who were not interviewed in Cohort 3. Participants were given two weeks to respond, and a reminder email was sent to non-responders approximately a week before the completion deadline. Table \ref{tab:01} presents information about the data collection method used in each cohort and the number (and percentage) of students from whom we collected data.  Interviews were also conducted with the faculty advisors after Cohort 1 and a subset of advisors after Cohort 3. Relevant data from these interviews is presented in the Discussion section to help interpret and elaborate on the results.

\begin{table}[H]
\centering
\caption{Number of participants (and percentages) from each DSC-WAV team that responded to the end-of-project survey or were interviewed.}
\label{tab:01}
\begin{tabular}{lll}
  \toprule
   & \textbf{Survey} & \textbf{Interview} \\
  \textbf{Team} & $N$ (\%) & $N$ (\%) \\
  \midrule
  \textit{Cohort 1 (Spring 2020)} & & \\[1ex]
    ~~Team 01 ($N=5$) & --- & 4 (80\%) \\
    ~~Team 02 ($N=5$) & --- & 4 (80\%) \\
    ~~Team 03 ($N=5$) & --- & 3 (60\%) \\
    ~~Team 04 ($N=5$) & --- & 5 (100\%)$^{\dagger}$ \\
    ~~Team 05 ($N=5$) & --- & 5 (100\%) \\
  \midrule
  \textit{Cohort 2 (Fall 2020} & & \\[1ex]
    ~~Team 06 ($N=4$) & 4 (100\%) & ---  \\
    ~~Team 07 ($N=3$) & 1 (33\%)  & ---  \\
    ~~Team 08 ($N=4$) & 3 (75\%)  & ---  \\
    ~~Team 09 ($N=3$) & 1 (33\%)  & ---  \\
    ~~Team 10 ($N=8$)$^{\dagger\dagger}$ & 4 (50\%) & --- \\
    ~~Team 11 ($N=2$) & 0 (0\%) & --- \\
  \midrule
  \textit{Cohort 3 (Spring 2021)} & & \\[1ex]
    ~~Team 12 ($N=5$) & 1 (20\%) & 3 (60\%)  \\
    ~~Team 13 ($N=5$) & 0 (0\%)  & 3 (60\%)  \\
    ~~Team 14 ($N=5$) & 0 (0\%)  & ---  \\
    ~~Team 15 ($N=4$) & 3 (75\%) & ---  \\
 \bottomrule
 \multicolumn{3}{l}{$^\dagger$One participant emailed their interview responses.}\\
 \multicolumn{3}{l}{$^{\dagger\dagger}$Three participants dropped out mid-project.}
\end{tabular}
\end{table}

Of note, the COVID-19 pandemic began during the middle of the project implementation for the first cohort. All of the teams, following institutional guidance, switched to remote settings mid-project. Because the program included multiple institutions, and each of these institutions took a different approach to the 2020--21 academic year, the second and third cohorts were primarily remote, but the academic schedules varied widely. This complicated the evaluation effort, resulting in differences in the evaluation actually undertaken across cohorts.

Although the items on the end-of-project survey and interviews were similar from cohort-to-cohort, the instruments were adapted after each cohort to facilitate better information and more nuance as the project evolved. For example, the survey for Cohort 2 incorporated items related to how the team members collaborated to complete the project. New items were also included to have student participants elaborate on their teams’ implementation of user stories, Kanban boards, code review, and use of GitHub. The survey and interview protocol for Cohort 3 contained similar open-ended questions inquiring about the students’ experience working on the project, their use of the adopted Agile framework, code review, GitHub, and any non-technical skills they learned working on the project. The faculty interviews centered around the same content, but aimed to gain insight into the faculty role in guiding their team throughout the project. Questions were asked to discern the level of faculty involvement needed from the students, how faculty perceived student growth throughout the project, and any feedback they wished to share about their experiences.

All three cohorts' student participant evaluation data  were examined and coded based on the following three themes: (1) implementation of the adopted Agile framework, (2) technical skills gained (e.g. GitHub, code review), and (3) non-technical skills gained (e.g. communication, collaboration, project management). Within each of these themes, the responses were summarized to answer the two research questions. The results of this synthesis are presented in the next section.

\section{Results}

Within this section, we present evidence from the synthesis of the evaluation data to provide answers to the two primary research questions. This evidence is presented longitudinally, by cohort, for each of the two research questions. The results from Cohorts 2 and 3 are combined since the evidence from the evaluation suggested similar student experiences and outcomes.

\subsection{How do undergraduate project teams actually implement Agile as they engage in a team-based data science project?}

\subsubsection{Cohort 1}

Although Cohort 1 was introduced to the modified Agile framework, students' responses from the end-of-project interviews suggested that the five teams did not implement the framework in a consistent manner. For example, the participants from three teams reported they regularly defined their work by using sprints and holding two to three weekly stand up meetings which discussed progress and future goals. One team that did not report using any part of the Agile framework held weekly meetings where tasks were divided and worked on during the meeting. The other team that reported not using the Agile framework had difficulties getting data and a research question from their community partner.  One student from that team explained that they tried hosting 15-minute stand ups during the week, but the meetings did not feel long enough to address all their issues. Eventually their team adopted a system of dividing all the work and then meeting later in the week for an hour to discuss their progress.

While only one team adopted the more formal sprint process, teams did adopt some of the other aspects of the modified Agile process. Two of the teams reported they used Kanban boards to organize their work, while the remainder relied on word-processing tools (e.g., Google Docs) to keep track of the team's backlog of tasks. All teams reported that they held a regular team meeting, but the frequency of this meeting varied across teams from 1--3 times a week. For many teams, this meeting did not resemble the shorter duration stand-ups that an Agile-based team would use, but instead teams held longer meetings fewer times a week in which they decided on next steps for the project and divided up the work. No teams reported holding any meetings resembling a sprint retrospective where they would reflect on what was working for their team and what was not. All teams reported meeting at least once with a community partner for a sprint demo.

\subsubsection{Cohorts 2 \& 3}

Results from Cohort 2’s and 3’s end-of-project survey responses indicated most teams seemed to be implementing the Agile framework more thoroughly than Cohort 1, employing elements of sprint planning, demos, and retrospectives. The student participants from all teams that completed the survey described planning sprints (about two weeks in duration) and using stand-up meetings, which occurred approximately three times a week, to monitor team progress. Students typically described their sprint planning meetings as ``team-driven events'' in which they set reasonable goals based on the “next logical step” for their project.

One main focus of the onboarding workshops for the second and third cohorts of students was facilitating work by mapping each task in the backlog to a user story (organized in a GitHub issue), which could be recorded and tracked on the team’s Kanban project board. The degree to which teams implemented this varied widely. Some teams were committed to this workflow, creating every task in the form of user stories, while other teams used aspects of the user stories in their backlog, and a few teams did not implement user stories at all. The teams that did adopt user stories seemed to appreciate how the stories helped make the tasks less abstract. Their user stories generally followed a similar format; describing the task, explaining its importance, and determining an endpoint. For example, one team member gave an example of this format, saying, \textit{``as a web app developer we would like to do this task for this reason and it will be completed when this code is in the repo.''}

Teams that partially adopted this framework also expressed how it helped them specify when a task was completed. One team indicated they would create a list of tasks in the backlog and add \textit{``I will know this task is complete when \ldots''} to each item. A few of the teams indicated that they did not specifically write user stories as they had learned about in the onboarding workshop. Some of these team members interviewed  said they forgot about user stories when working on the project, while other students explained that they did not feel like the user stories were efficient or useful for their project. These teams, however, often did still keep track of their to-do list on the Kanban board or backlog.

All Cohort 2 and 3 respondents indicated that their teams performed some sort of sprint retrospective. How this retrospective was conducted also varied across teams, with some teams incorporating a specific retrospective meeting and others engaging in reflection as they planned the subsequent sprint. Regardless of format, the students indicated they used this time to reflect on what was going well with the project, what areas needed more work, and to give each other advice.

In the next sections, we provide an answer to the second research question: to what extent does engaging in a team-based data science project help students develop technical and non-technical skills? The evidence is again presented longitudinally by cohort. We address the development of students' technical and nontechnical skills separately.

\subsection{To what extent does engaging in a team-based data science project help students develop technical skills?}

\subsubsection{Cohort 1}

Students in Cohort 1 described learning a variety of technical skills while working on their projects. Notably, students reported learning technical skills they felt they would not have had a chance to learn at the same depth in the courses offered at their institutions. These skills included coding in R and Python, statistical analysis applications, working with relational databases, reading R documentation, and creating complex visualizations and maps.

As students’ individual technical proficiencies developed throughout the project, they often shared their new knowledge with their teammates. One student noted that, though each member of the group had a different strength, they often worked together to share ideas, resources, and teach their teammates to code. A student on a different team explained that they were able to help students with less coding experience than they. The student said, \textit{``[they] \ldots didn't necessarily mentor the other students in the group, but explained some lines of code to other students if they asked \ldots . This was good as a review for me and to share.''} Because students completed their projects in a group they were able to grow and share their technical skills while working on the project in a way they would not have if they had to work completely independently on the project.

Almost all of the students also indicated that the project helped them to learn and become more comfortable using Git and GitHub. For many student participants, this project was their first experience using Git and GitHub. One student noted that although it was intimidating, especially in the beginning, it was okay since the team was learning GitHub together. Most students also came to appreciate the advantages of using a version control system for team-based data science, with one student describing how it was, \textit{``good to see collaborative atmosphere in github and all use what other people wrote to push [the project] to the next level.''}

Although all five teams reported using GitHub for their projects, the way they used GitHub varied from project to project. Some teams consistently committed updates to their repositories, while others only uploaded materials they considered to be in a final state. One student confessed that they were nervous to make commits because they were, \textit{``concerned that the code wasn't refined enough or `prime-time' ready.''} This is something the faculty made note of as well, and tried during future cohorts to clarify the ways in which the ungraded collaborative project differed from work students might be doing for a course. The students typically did not commit their work as they progressed, but they would upload it just in time for meetings. Seeing as that is not typically what is done in practice for data science (which often exhibits regular pull/change/push cycles), the project leadership prioritized encouraging students to commit more often for the next cohort of students.

In Cohort 1, developing code review skills was not yet a focus of DSC-WAV. However, during an interview a student explained that they wished there had been coding standards and styles outlined at the beginning of the project to set expectations for how work should be completed on the team. This led to discussions that prompted the faculty to promote code review in future cohorts.

\subsubsection{Cohorts 2 \& 3}

Students in Cohorts 2 and 3 described a large number of technical data science skills they acquired from working on their project. They specifically alluded to developing skills related to text processing, working with HTML files, tidying data, visualizing data, writing functions and coding in Python and R, and creating dashboards via Shiny Apps. Students from every team reported gaining more experience using Git and GitHub, using these tools to facilitate work on the project. When asked how the teams used GitHub, students described using it for file and code storage, as well as for tracking their project progress through the Kanban board. One team also made use of the GitHub Wiki feature to record notes and questions they had throughout the project.

Almost every student interviewed indicated they felt more comfortable using GitHub after the project. Students largely reported being more confident making commits, pulling/pushing updates, solving merge conflicts, branching, creating issues, and using the Kanban board feature. One student summing this up stated, \textit{``my GitHub skills started off quite rusty, and I feel like they have improved exponentially, and I now feel more confident collaborating with GitHub.''} While this improved confidence seemed reflective of many students' experiences, there were still aspects of the GitHub process that students expressed uncertainty about. For example, even after using it for a semester, one student indicated she still felt uncomfortable using the terminal to update the repository and dealing with more complex merge issues or error messages. The students who expressed frustration that they had to learn Git and GitHub as they worked on the project, said they would have preferred to have more GitHub resources or training before beginning the project.

Most teams in Cohorts 2 and 3 also seemed to engage in some form of code review, although this process varied widely across teams. One team, for example, had a Scrum leader that was the only person who reviewed the code. Another team described a cyclic system of team members reviewing each other's code. Another team implemented a partner system for reviewing the code, and another reviewed all the code together as a group. There were some teams from these cohorts that did not implement code review. These teams often had projects that did not require as much coding, or the code they did have was viewed as not complex enough to warrant a code review. For example, one student explained that their team:

\begin{quote}
``reviewed the code amongst [them]selves, but did not formally do a code review. Part of it was because the project did not necessitate complex code, but also since any code we made was to convert or alter weather/water data, [they] could immediately see whether it was working or not.''
\end{quote}

Although they had a process in place for who would carry out the code review, most teams reported that the review process itself was fairly informal, checking only to evaluate that the code being reviewed was functional or that there were no merge conflicts introduced. Most teams did not tend to examine their code for style or consistency. Members from two teams did mention that they reviewed their code for style, comprehensibility, and comments but could not specify what criteria were actually being used for ensuring those features were up to team standards. Another team (from the third cohort) evaluated their code for consistency and style by having a single new team member go through the team’s code from the previous semester to clean it up and add comments. This process seemed to help onboard this student and ensure the team’s code was well commented to pass off to the community partner.

\subsection{To what extent does engaging in a team-based data science project help students develop non-technical skills?}

\subsubsection{Cohort 1}

The interviews with Cohort 1 students focused on the students' perceived growth in non-technical skills that were gained by working on a team-based data science project. Students mentioned experiences with communication, collaboration, and project management that led to growth for them as team members and leaders.

Many students indicated that their teamwork and communication skills improved while working on the project. Several interviewees referred to how the team-based approach helped them learn how to communicate with other team members and listen to their perspectives, even changing their perception of team-based work. One student said:

\begin{quote}
``\ldots a lot of people don't like group projects, but I've really come to appreciate working in a group. This group got along really well. We were doing the right things to make a group work: communicating well, dividing work fairly, setting reasonable goals both individually and for the group.''
\end{quote}

Several students explained that even though they worked in a group they needed to develop the confidence and initiative needed to further their part of the project or to facilitate the work among their team members. One student working as the community engagement scholar for their team not only took on a leadership role, but also encouraged teammates to learn additional skills beyond the technical aspects of the project:

\begin{quote}
``I already had a strong leadership background, but I did improve and strengthen over time. At the beginning we had met with the person at [company name removed] and then we were told to learn more soft skills. From this I learned that I needed to take more of a leadership role and teach teammates the skills of talking with the business side as well. I encouraged group mates to talk to the client as well to get experience explaining the work to the client. Glad to be placed in this role, though I wanted to do data science this made me more confident talking to people outside the domain.''
\end{quote}

Two groups described communication and teamwork issues among group members. Sometimes students did not communicate enough, only to find they were working on the same parts of the project in a duplicative fashion. A student pointed out that if there were more of a framework for the workflow in place in their group they might not have had this issue. The student's reflection, along with many other students' comments on group work, led to including more non-technical skills for collaboration into the onboarding workshop for the second and third cohorts.

\subsubsection{Cohorts 2 \& 3}

The majority of the students in these cohorts also highlighted non-technical skills (communication, teamwork, and project management) when asked about the proficiencies they thought the project helped them develop. One student explained that they \textit{``gained experience working with a group on a product for a client and in communicating directly with the client to meet their expectations.''} While another student pointed out that they feel much more confident \textit{``breaking down large tasks into smaller, more manageable ones.''} Though the students were not directly asked about project management and communication in the survey,  the students provided great detail about their process for sprint planning, coordinating work, and project management through their use of Kanban boards. Working on this team-based data science project using the Agile framework gave the students the chance to develop the professional communication skills needed to complete the task.

Students in these cohorts also reported that their teams engaged much more frequently with the community partner than the teams in Cohort 1. Moreover, while this engagement in Cohort 1 was primarily undertaken by the community engagement scholar on the team, in Cohorts 2 and 3 the other team members also engaged with the community partner, presenting their work in a variety of formats ranging from presentations, posters, verbal reports, and written reports. The frequency and degree of this engagement varied across teams, with some teams meeting only once with the community partner to review the semester’s progress, and others presenting this progress after each sprint and even engaging the community partner in their sprint planning and team meetings.

Engaging with the community partner was identified as a crucial part of some projects. One student alluded to this, saying, \textit{``our client was essential in filling in the gaps, explaining how the data was collected and what she hoped we would accomplish as a team.''} For the teams that had less engagement with the community partner, this seemed to be more a function of the community organization than the student team. One such team reported that their community partner was not very responsive, so they did not meet with them that often. Another, citing the community partner's busy schedule, indicated they only met or emailed with the community partner when that person had availability.

Lastly, the students in these cohorts predominantly named their team members, more than faculty mentors or community partners, as the force driving the project in terms of direction, setting meetings, and holding members accountable for their goals. While acknowledging that their faculty mentors were incredibly helpful, several pointed out that in these cohorts the faculty mentors were more hands-off and allowed the students to have ownership of the project. One student made the following statement about their project experience:

\begin{quote}
``It felt like a simulation of what industry is like as opposed to another group project for class. We didn’t have a professor assigning us tasks. Rather we as a team had to discuss the best way to move forward with the project.''
\end{quote}

This independence not only allowed students to develop leadership and project management skills, but granted them space to create their own supportive team culture. Two teams in Cohort 3 described creating open environments for their meetings in which they could confide in each other, \textit{``have each other’s backs,''} freely ask questions, and learn together. One student summed up their experience with the following description of their team meeting atmosphere:

\begin{quote}
``We were all on the same team and so we became friends through the process. We all wanted to make the best thing possible we could. So we were really willing to help each other out.''
\end{quote}

\section{Discussion}

The collaborative nature of data science projects in the workplace make experiences with team-based data science important for data science and statistics students. Team-based data science skills, a critical component of data acumen, are not often taught in undergraduate statistics courses in which problems are typically well-defined. The DSC-WAV program promoted team-based data science skills by engaging students in authentic data analysis for a community organization. The program employed an Agile framework to help students organize and carry out the project.  Within this research, we sought to understand how the DSC-WAV teams enacted the Agile framework as they engaged in team-based data science and the extent to which this framework supported development of students' technical and non-technical data science skills.

Overall, teams in all three cohorts demonstrated the ability to implement versions of the Agile framework that helped facilitate the completion of their projects. Cohort 1 had more variability in implementing the Agile framework than later cohorts. This cohort typically had fewer team meetings and typically spent their team meeting time dividing tasks among the students. Teams in Cohorts 2 and 3 reported more frequent team meetings that were more team driven and included more sprint retrospective meetings and sprint demos than previous cohorts. The later cohorts also reported more use of the Kanban boards and user stories to organize team work, likely due to the increased focus of the onboarding workshops on these topics.

The process of determining and maintaining this system was helped by having faculty mentors who early in the project strongly emphasized the importance of frequent team meetings. In teams that reported having at least three meetings a week and planning regular sprints, the students reported feeling more ownership of the project. These students often described planning and running their own meetings, only needing the faculty advisor once a week to consult on the project. The Agile structure helped the students feel more autonomous working on the project as a team, rather than relying on the faculty mentor to lead the group at every meeting. This autonomy speaks to the potential sustainability of this type of project. We observed, anecdotally, that if faculty mentors invest time early in the term to get students working in an Agile framework, teams tend to take on more ownership and become more autonomous and effective as they get more experience.

Students' perceptions of their technical and non-technical skills were improved after taking part in this program. Students reported more confidence in their programming, visualization, and data analytic skills. They also reported greater confidence using GitHub for a team project. Many students said they had never used GitHub before, but came out of the project feeling more confident working with it. In comparison to the first cohort, Cohorts 2 and 3 teams more fully incorporated GitHub into their project workflow (possibly as a result of increased focus on GitHub during the onboarding workshop). Most teams in Cohorts 2 and 3 also experienced some version of a code review. Though the code review was typically checking whether the code was functional, this is still an experience not common in undergraduate courses in statistics and data science.

Across cohorts, many students also indicated that working on the DSC-WAV program helped them improve their teamwork and communication skills. They specifically identified their communication with other team members and with the client as improving because of the programs’s organizational structure. Although the frequency of communication with the community partner improved in Cohorts 2 and 3 after addressing it in the onboarding for these cohorts, it was not frequent enough to provide a constant cycle of iterative feedback from the stakeholders. Several students also noted that the program helped improve their perception of team-based work, many contrasting this experience with group-based work experienced in their coursework.

Although many positive themes emerged from the students' responses to the end of term surveys and interviews, there were many challenges that came with orchestrating a program of this nature.

\subsection{Limitations, Challenges, and Lessons Learned}

The study had several limitations. The response rates of surveys decreased noticeably from 80\% for Cohort 1 to slightly over 50\% in Cohort 3 (see Table \ref{tab:01}). While this is not surprising given the exhaustion and fatigue students experienced because of the pandemic and remote work, the findings have to be considered in light of this attrition. Furthermore, we only have access to students' perceptions of their growth in their skillset, and not an actual measure of their actual growth. This is in large part because instruments to measure competency in data science do not exist (or at least were not mature enough to be used in a study).

All cohorts reported feeling the effects of remote work during the COVID-19 pandemic. This was especially true for the first cohort, who went remote halfway through the semester. Many teams in this cohort felt they did not have the same quality of work or sense of community after they went online. The students had difficulty finding the time to meet synchronously and keeping up communication. Students in the second and third cohorts did not indicate that the remote setting impacted their work quality nor their sense of community. In fact, some students in Cohorts 2 and 3 felt they thrived holding stand-ups online, participating in remote working meetings, and sharing their screens to facilitate writing code. With new remote working and learning skills it may be worthwhile to consider encouraging use of remote work options for future cohorts. For example, using Zoom for remote stand-up meetings will make it much easier for teams to schedule these short sessions, and Slack with Zoom integration can help them plan and shift easily into synchronous meetings. During Cohort 1, one team had students from multiple campuses and utilized remote meeting options for the stand-up meetings each week, demonstrating that this was a successful collaboration method.

Despite the team-based approach to data science promoted by the DSC-WAV program, the five teams from Cohort 1 largely completed tasks using a ``divide-and-conquer'' approach to the work in which individuals would take on different tasks. Two teams reported that, although they divided the work, they did occasionally meet as a larger team to share what they learned and resources they found while working on their part of the project. The divide-and-conquer approach to work is reminiscent of most student ``collaborations'' in the classroom and does not reflect how team-based work is done in the workplace. It is unclear whether this approach was taken by the students because it is what they are used to doing, or because the projects being undertaken did not require a collaborative approach. Moreover, many students stated that working remotely affected their abilities to meet regularly and organize the work. This may also have impacted how they collaborated on the project.

In general, the teams in Cohorts 2 and 3 reported better fidelity to the adopted Agile framework than did the teams in Cohort 1. This might be due, in part, to the increased focus on the Agile process during the cohorts' onboarding, as well as more promotion of the process from the faculty advisors. With each new cohort, the faculty mentors learned more about how to manage these projects. Many faculty reported in interviews at the end of Cohort 1 that they did not know the level of involvement they should take in directing their team's meetings. The evaluation reports containing summaries of student reflections allowed faculty mentors to adjust their implementation and better meet the needs of the students. Faculty interviewed at the end of Cohort 3 (including faculty mentors who are co-authors on this paper) indicated that they took a more hands-on role modeling the Agile framework and facilitating students' use of GitHub to get their teams started on the project for Cohorts 2 and 3, but stepped back a bit later in the semester, when the students took more of a leading role.  Considering the positive feedback from faculty mentors and students, this system worked well. One of the Cohort 3 faculty members interviewed indicated that using the Agile structure has allowed them to \textit{``spend less time with [the students] than [they] have in the past and be more productive with what my contributions [were].''} Additionally, they noted that their students were much more comfortable with GitHub in the third cohort, only needing help a few times throughout the semester.

While the community engagement scholar was seen as a valuable contributor on most teams, the role of this student could be more clearly defined in future cohorts. These students often modeled the professional and communication skills needed to work with the community organizations and run team meetings. Although these are team-based data science projects, they need a student leader to facilitate meetings and drive the project forward. The community engagement scholar was well suited for this role. They served as the liaisons between the data science students and the community partner organizations. Often they spent time researching and communicating with the organizations to determine the path of the project. Then they conveyed that information to the data science students.

A few community engagement scholars in each cohort reported that they were uncertain about their role on the project. One particular student described how they got strong signals from the onboarding workshop and their faculty mentor that the project should be directed by students as a team. This led students to resist taking the lead role in meetings for fear of losing a sense of team ownership of the project. Ultimately, the lack of a clear leader wasted some time in meetings for this particular team. In response to this, the community engagement scholar suggested that their role be more well-defined for future cohorts. They advocated for a document that details specific tasks and the nature of the position so future students can get an idea of how they can best take on this role. Minimally, this student should take the role of leading meetings to help the project progress, not dictating the direction but engaging the data science students in discussion of potential directions.

Creating and delivering this experience to students took a great deal of time from the faculty advisors, a resource that is scarce for most academic faculty. To begin, the advisor needed to build a relationship with the community organizations, which was key to gaining their trust and developing the artifacts that the organizations were looking for. The advisors also needed to meet regularly with their teams during the project to help them implement Agile, provide resources and feedback, guide them through the cycle of inquiry and analysis, and act as cheerleader. This degree of faculty involvement leads to questions of how feasible it is to scale this project for future cohorts of students.

Finally, we note a single semester may not be adequate for this type of team-based project. In proposing the program, we considered a number of models, including semester-long, year-long, and summer projects. We decided to set the duration of the project to one-semester to enlarge the participant pool and the number of students impacted by the program. This would also have the advantage of program contact with more community partners which would provide more examples of the applicability of data science. However, after the first cohort, we felt that one semester may not be long enough to allow students to carry out the more in-depth work that the community partners would benefit from. Subsequently, we extended some of our more successful one-semester projects to carry over into additional semesters. For some teams, the work continued with the same team of students from the previous cohort, but for others, only a subset of the previous cohorts’ students continued and some new students were onboarded and added to the team.

\subsection{Conclusion}

The inspiration for the DSC-WAV program was a question of whether undergraduate students could tackle real-world data science problems utilizing the tools and approaches frequently seen in industry. Based on our experiences, the answer to this question is ``yes.'' We believe we were able to provide undergraduate students an authentic team-based experience working with more complex data and on more challenging problems than they had experienced in their coursework. We also believe that these experiences allowed them to develop some of the technical and non-technical skills that are valued in the workplace.

Carrying out this program with three cohorts of students has given us many insights into the feasibility of implementing team-based data science projects at the undergraduate level. With each cohort, the program's leadership team and faculty advisors learned and made refinements to (we believe) improve the tools and processes for subsequent implementations. This experience and our reflections have given us insight about how team-based data science can be implemented in similar projects.

The DSC-WAV program effectively partners with community organizations to help them with their data analysis projects. This type of experience might have similar benefits for students looking to enter the workforce, as REUs do for students who want to pursue academia, or go to graduate school. While this program utilized teams working during the semester, similar approaches could be implemented in courses, through other co-curricular experiences, or during summer programs.

Our experience with the DSC-WAV program also points to potential tools and processes that can be adopted within the academic curriculum. For example, data science coursework could provide students with more team-based data science experiences such as using GitHub, participating in code review, working with real data to solve ill-structured problems, and communicating results. Moreover, these types of experiences can also be provided to students early in their undergraduate careers. As we have seen, this type of work is not beyond their reach after completing only a few data science courses.

The literature and current state of data science suggests that team-based data science skills are in demand and will continue to be for some time \cite{NASEM:2018}. Though there is still much work to be done to determine effective ways of introducing students to data science in authentic ways, we believe that the DSC-WAV program has allowed us to make progress toward this longer-term goal.

\section*{Acknowledgments}

We acknowledge the support of NSF grants HDR DSC-1923388, HDR DSC-1923700, HDR DSC-1923934, and HDR DSC-1924017. We also want to acknowledge the community organizations that participated in the project and Andrea Dustin, whose organization and faculty-herding skills have kept this project moving forward.

\appendix

\section{Description of the Onboarding Workshops}

Each cohort participated in an onboarding workshop to help orient them to the DSC-WAV program and their particular project. These workshops also addressed specific technical (e.g., Git/GitHub, code review) and non-technical (e.g., team-building, communication) skills. With the exception of Cohort 3, which was done asynchronously, the other onboardings have been conducted synchronously during two half-day remote workshops (Cohort 2) or during a full-day in-person (Cohorts 1 and 4) workshop.

While the specific content for the onboarding process has evolved over the project period, it currently includes content related to the Agile process (Scrum roles, sprint mechanics, the use of Kanban project boards, writing effective user stories), using Git and GitHub (project boards, pull requests, handling git conflicts), and undertaking code review. Starting with Cohort 2, a data science professional from Atlassian has also led an interactive demonstration (during the onboarding workshop) to facilitate the writing of user stories. Details of each cohort's onboarding workshop can be found at the following URL: \href{https://dsc-wav.github.io/www/outreach.html}{https://dsc-wav.github.io/www/outreach.html}.

Prior to the onboarding workshop, students are asked to complete several tasks including familiarizing themselves with the DSC-WAV program code of conduct, joining the team's Slack channels and working through some Slack-related tasks, installing R, RStudio and Git, creating a GitHub account, cloning the project's repository and linking it to RStudio, and completing the institution required CITI training for the responsible conduct of research. Expectations related to the commitment of regularly meeting with their team and faculty coordinator are also set so that students are able to put these meetings on their calendar during the onboarding workshop. (Teams are required to meet no fewer than three times per week. This requirement was instituted after the lack of shared meeting times made it difficult for some teams from previous cohorts to effectively engage.) Details of the pre-onboarding tasks can be found at: \href{https://dsc-wav.github.io/www/preonboarding.html}{https://dsc-wav.github.io/www/preonboarding.html}.


\bibliographystyle{AIMS}
\bibliography{fds-2021}

\providecommand{\href}[2]{#2}
\providecommand{\arxiv}[1]{\href{http://arxiv.org/abs/#1}{arXiv:#1}}
\providecommand{\url}[1]{\texttt{#1}}
\providecommand{\urlprefix}{URL }
\begin{thebibliography}{10}

\bibitem{Amason:1997}
\newblock A.~Amason and H.~Sapienza,
\newblock The effects of top management team size and interation norms on
  cognitive and affective conflict,
\newblock \emph{Journal of Management}, \textbf{23} (1997), 495--516,
\newblock \urlprefix\url{https://doi.org/10.1177/014920639702300401}.

\bibitem{Baumer:2015}
\newblock B.~Baumer,
\newblock A data science course for undergraduates: Thinking with data,
\newblock \emph{The American Statistician}, \textbf{69} (2015), 334--342,
\newblock \urlprefix\url{https://doi.org/10.1080/00031305.2015.1081105}.

\bibitem{Beckman:2021}
\newblock M.~D. Beckman, M.~{{\c{C}}etinkaya{-}Rundel}, N.~J. Horton, C.~W.
  Rundel, A.~J. Sullivan and M.~Tackett,
\newblock Implementing version control with {Git} and {GitHub} as a learning
  objective in statistics and data science courses,
\newblock \emph{Journal of Statistics and Data Science Education}, \textbf{29}
  (2021), 1--35,
\newblock \urlprefix\url{https://doi.org/10.1080/10691898.2020.1848485}.

\bibitem{Bryan:2018}
\newblock J.~Bryan,
\newblock Excuse me, do you have a moment to talk about version control?,
\newblock \emph{The American Statistician}, \textbf{72} (2018), 20--27,
\newblock \urlprefix\url{https://doi.org/10.1080/00031305.2017.1399928}.

\bibitem{Bureau-of-Labor-Statistics:2021}
\newblock {Bureau of Labor Statistics} and {U.S. Department of Labor},
\newblock Occupational outlook handbook: Computer and information research
  scientists,
\newblock Online, 2021,
\newblock
  \urlprefix\url{https://www.bls.gov/ooh/computer-and-information-technology/computer-and-information-research-scientists.htm}.

\bibitem{Burke:2006}
\newblock C.~S. Burke, K.~C. Stagl, E.~Salas, L.~Pierce and D.~Kendall,
\newblock Understanding team adaptation: A conceptual analysis and model,
\newblock \emph{Journal of Applied Psychology}, \textbf{91} (2006), 1189--1207,
\newblock \urlprefix\url{http://dx.doi.org/10.1037/ 0021-9010.91.6.1189}.

\bibitem{Busjahn:2011}
\newblock T.~Busjahn, C.~Schulte and A.~Busjahn,
\newblock Analysis of code reading to gain more insight in program
  comprehension,
\newblock in \emph{{Koli Calling '11: Proceedings of the 11th Koli Calling
  International Conference on Computing Education Research}}, 2011,
\newblock 1--9,
\newblock \urlprefix\url{https://doi.org/10.1145/2094131.2094133}.

\bibitem{Cetinkaya-Rundel:2021}
\newblock M.~{{\c{C}}etinkaya{-}Rundel} and V.~Ellison,
\newblock A fresh look at introductory data science,
\newblock \emph{Journal of Statistics and Data Science Education}, \textbf{29}
  (2021), S16--S26,
\newblock \urlprefix\url{https://doi.org/10.1080/10691898.2020.1804497}.

\bibitem{Chow:2008}
\newblock T.~Chow and D.-B. Cao,
\newblock A survey of critical success factors in {A}gile software projects,
\newblock \emph{Journal of Systems and Software}, \textbf{81} (2008), 961--971,
\newblock \urlprefix\url{https://doi.org/10.1016/j.jss.2007.08.020}.

\bibitem{Condon:2021}
\newblock S.~Condon,
\newblock Data science is a team sport: How to choose the right players,
\newblock {ZD Net}, 2021,
\newblock
  \urlprefix\url{https://www.zdnet.com/article/data-science-is-a-team-sport-how-to-choose-the-right-players/}.

\bibitem{Conforto:2014}
\newblock E.~C. Conforto, F.~Salum, D.~C. Amaral, S.~L. da~Silva and L.~F.~M.
  de~Almeida,
\newblock Can agile project management be adopted by industries other than
  software development?,
\newblock \emph{Project Management Journal}, \textbf{45} (2014), 21--34,
\newblock \urlprefix\url{https://doi.org/10.1002/pmj.21410}.

\bibitem{Dillman:2014}
\newblock D.~A. Dillman, J.~D. Smyth and L.~M. Christian,
\newblock \emph{Internet, Mail, and Mixed-Mode Surveys: The Tailored Design
  Method},
\newblock 4th edition,
\newblock Wiley, New York, 2014.

\bibitem{Donoghue:2021}
\newblock T.~Donoghue, B.~Voytek and S.~E. Ellis,
\newblock Teaching creative and practical data science at scale,
\newblock \emph{Journal of Statistics and Data Science Education}, \textbf{29}
  (2021), 1--22,
\newblock \urlprefix\url{https://doi.org/10.1080/10691898.2020.1860725}.

\bibitem{Eghbal:2020}
\newblock N.~Eghbal,
\newblock \emph{Working in Public: The Making and Maintenance of Open Source
  Software},
\newblock Stripe Press, San Francisco, CA, 2020.

\bibitem{Fisher:1997}
\newblock S.~G. Fisher, T.~A. Hunter and W.~D.~K. Macrosson,
\newblock Team or group? managers' perceptions of the differences,
\newblock \emph{Journal of Managerial Psychology}, \textbf{12} (1997),
  232--242,
\newblock \urlprefix\url{https://doi.org/10.1108/02683949710174838}.

\bibitem{Forsgren:2018}
\newblock E.~Forsgren and K.~Bystr{\"o}m,
\newblock Multiple social media in the workplace: Contradictions and
  congruencies,
\newblock \emph{Information Systems Journal}, \textbf{28} (2018), 442--464,
\newblock \urlprefix\url{https://doi.org/10.1111/isj.12156}.

\bibitem{Fredrick:2008}
\newblock T.~A. Fredrick,
\newblock Facilitating better teamwork: Analyzing the challenges and strategies
  of classroom-based collaboration,
\newblock \emph{Business Communication Quarterly}, \textbf{71} (2008),
  439--455,
\newblock \urlprefix\url{https://doi.org/10.1177/1080569908325860}.

\bibitem{Goleman:1998}
\newblock D.~Goleman,
\newblock What makes a leader?,
\newblock \emph{Harvard Business Review}, \textbf{76} (1998), 93--102,
\newblock \urlprefix\url{https://hbr.org/2004/01/what-makes-a-leader}.

\bibitem{Hardin:2015}
\newblock J.~Hardin, R.~Hoerl, N.~J. Horton, D.~Nolan, B.~Baumer, O.~Hall-Holt,
  P.~Murrell, R.~Peng, P.~Roback, D.~T. Lang and M.~D. Ward,
\newblock Data science in statistics curricula: Preparing students to ``think
  with data'',
\newblock \emph{The American Statistician}, \textbf{69} (2015), 343--353,
\newblock \urlprefix\url{https://doi.org/10.1080/00031305.2015.1077729}.

\bibitem{Harris:1996}
\newblock P.~R. Harris and K.~G. Harris,
\newblock Managing effectively through teams,
\newblock \emph{Team Performance Management: An International Journal},
  \textbf{2} (1996), 23--36,
\newblock \urlprefix\url{https://doi.org/10.1108/13527599610126247}.

\bibitem{Hoda:2011}
\newblock R.~Hoda, J.~Noble and S.~Marshall,
\newblock The impact of inadequate customer collaboration on self-organizing
  {A}gile teams,
\newblock \emph{Information and Software Technology}, \textbf{53} (2011),
  521--534,
\newblock \urlprefix\url{https://doi.org/10.1016/j.infsof.2010.10.009}.

\bibitem{Hossain:2009}
\newblock E.~Hossain, M.~A. Babar and H.~Paik,
\newblock Using scrum in global software development: A systematic literature
  review,
\newblock in \emph{2009 Fourth {IEEE International Conference on Global
  Software Engineering}}, 2009,
\newblock 175--184,
\newblock \urlprefix\url{https://doi.org/10.1109/ICGSE.2009.25}.

\bibitem{Hsing:2019}
\newblock C.~Hsing and V.~Gennarelli,
\newblock Using {GitHub} in the classroom predicts student learning outcomes
  and classroom experiences: Findings from a survey of students and teachers,
\newblock in \emph{{SIGCSE '19: Proceedings of the 50th ACM Technical Symposium
  on Computer Science Education}},
\newblock New York, 2019,
\newblock 672--678,
\newblock \urlprefix\url{https://doi.org/10.1145/3287324.3287460}.

\bibitem{Johnson:1995}
\newblock D.~W. Johnson and R.~T. Johnson,
\newblock Social interdependence---cooperative learning in education,
\newblock in \emph{Conflict, Cooperation, and Justice} (eds. B.~Bunker and
  J.~Z. Rubin),
\newblock Jossey-Bass Publishers, San Francisco, 1995,
\newblock 205--251.

\bibitem{Johnson:1999}
\newblock D.~W. Johnson and R.~T. Johnson,
\newblock \emph{Learning Together and Alone: Cooperative, Competitive, and
  Individualistic Learning},
\newblock 5th edition,
\newblock Allyn and Bacon, Needham Heights, Massachusetts, 1999.

\bibitem{Kastl:2018}
\newblock P.~Kastl and R.~Romeike,
\newblock Agile projects to foster cooperative learning in heterogeneous
  classes,
\newblock in \emph{{2018 IEEE Global Engineering Education Conference
  (EDUCON)}}, 2018,
\newblock 1182--1191,
\newblock \urlprefix\url{https://doi-org/10.1109/EDUCON.2018.8363364}.

\bibitem{Kim:2016}
\newblock M.~Kim, T.~Zimmermann, R.~DeLine and A.~Begel,
\newblock The emerging role of data scientists on software development teams,
\newblock in \emph{{ICSE} '16: Proceedings of the 38th {I}nternational
  {C}onference on {S}oftware {E}ngineering}, 2016,
\newblock 96--107,
\newblock \urlprefix\url{https://doi.org/10.1145/2884781.2884783}.

\bibitem{KnowledgeWharton:2019}
\newblock {Knowledge@Wharton},
\newblock What's driving the demand for data scientists?, 2019,
\newblock
  \urlprefix\url{https://knowledge.wharton.upenn.edu/article/whats-driving-demand-data-scientist/}.

\bibitem{Leach:2005}
\newblock D.~J. Leach, T.~D. Wall, S.~G. Rogelberg and P.~R. Jackson,
\newblock Team autonomy, performance, and member job strain: Uncovering the
  teamwork {KSA} link,
\newblock \emph{Applied Psychology}, \textbf{54} (2005), 1--24,
\newblock \urlprefix\url{https://doi.org/10.1111/j.1464-0597.2005.00193.x}.

\bibitem{Limoncelli:2020}
\newblock T.~A. Limoncelli,
\newblock Five nonobvious remote work techniques,
\newblock \emph{Communications of the {ACM}}, \textbf{63} (2020), 108--110,
\newblock
  \urlprefix\url{https://cacm.acm.org/magazines/2020/11/248217-five-nonobvious-remote-work-techniques/fulltext}.

\bibitem{Lopez-Alcarria:2019}
\newblock A.~L{\'o}pez-Alcarria, A.~Olivares-Vicente and F.~Poza-Vilches,
\newblock A systematic review of the use of agile methodologies in education to
  foster sustainability competencies,
\newblock \emph{Sustainability}, \textbf{11} (2019), 2915,
\newblock \urlprefix\url{https://doi.org/10.3390/su11102915}.

\bibitem{Luca:2001}
\newblock J.~Luca and P.~Tarricone,
\newblock Does emotional intelligence affect successful teamwork?,
\newblock in \emph{Meeting at the Crossroads: Proceedings of the 18th Annual
  Conference of the Australasian Society for Computers in Learning in Tertiary
  Education},
\newblock Melbourne, Australia, 2001,
\newblock \urlprefix\url{https://ro.ecu.edu.au/ecuworks/4834}.

\bibitem{Maier:2019}
\newblock T.~Maier, J.~DeFranco and C.~Mccomb,
\newblock An analysis of design process and performance in distributed data
  science teams,
\newblock \emph{Team Performance Management: An International Journal},
  \textbf{25} (2019), 419--439,
\newblock \urlprefix\url{https://doi.org/10.1108/TPM-03-2019-0024}.

\bibitem{Mannix:2005}
\newblock E.~Mannix and M.~A. Neale,
\newblock What difference makes a difference: The promise and reality of
  diverse groups in organizations,
\newblock \emph{Psychological Science in the Public Interest}, \textbf{6}
  (2005), 31--55,
\newblock \urlprefix\url{https://doi.org/10.1111/j.1529-1006.2005.00022.x}.

\bibitem{McNamara:2017}
\newblock A.~McNamara, N.~J. Horton and B.~S. Baumer,
\newblock Greater data science at baccalaureate institutions,
\newblock \emph{Journal of Computational and Graphical Statistics}, \textbf{26}
  (2017), 781--783,
\newblock \urlprefix\url{https://doi.org/10.1080/10618600.2017.1386568}.

\bibitem{NASEM:2018}
\newblock {National Academies of Science, Engineering, and Medicine},
\newblock \emph{Data Science for Undergraduates: Opportunities and Options},
\newblock The National Academies Press, Washington, DC, 2018,
\newblock \urlprefix\url{https://doi.org/10.17226/25104}.

\bibitem{Nolan:2010}
\newblock D.~A. Nolan and D.~{Temple Lang},
\newblock Computing in the statistics curriculum,
\newblock \emph{The American Statistician}, \textbf{64} (2010), 97--107,
\newblock \urlprefix\url{https://doi.org/10.1198/tast.2010.09132}.

\bibitem{Qumer:2008}
\newblock A.~Qumer and B.~Henderson-Sellers,
\newblock An evaluation of the degree of agility in six {A}gile methods and its
  applicability for method engineering,
\newblock \emph{Information and Software Technology}, \textbf{50} (2008),
  280--295,
\newblock \urlprefix\url{https://doi.org/10.1016/j.infsof.2007.02.002}.

\bibitem{Rawlings-Goss:2018}
\newblock R.~Rawlings-Goss, L.~Cassel, M.~Cragin, C.~Cramer, A.~Dingle,
  S.~Friday-Stroud, A.~Herron, N.~Horton, T.~R. Inniss, K.~Jordan,
  P.~Ord{\'o}{\~n}ez, R.~Rwebangira, K.~Schmitt, D.~Smith and S.~Stephens,
\newblock Keeping data science broad: Negotiating the digital and data divide
  among higher-education institutions,
\newblock Workshop: Bridging the Digital and Data Divide, National Science
  Foundation, 2018,
\newblock \urlprefix\url{https://par.nsf.gov/biblio/10075971}.

\bibitem{Razmov:2006}
\newblock V.~Razmov and R.~Anderson,
\newblock Experiences with agile teaching in project based courses,
\newblock in \emph{Proceedings of the American Society for Engineering
  Education, ASEE Annual Conference \& Exposition},
\newblock American Society for Engineering Education, Chicago, 2006,
\newblock 11.615.1--11.615.12,
\newblock \urlprefix\url{https://doi.org/10.18260/1-2--1018}.

\bibitem{Sadowski:2018}
\newblock C.~Sadowski, E.~S{\"o}derberg, L.~Church, M.~Sipko and A.~Bacchelli,
\newblock Modern code review: A case study at {G}oogle,
\newblock in \emph{{ICSE-SEIP '18: Proceedings of the 40th International
  Conference on Software Engineering: Software Engineering in Practice}}, 2018,
\newblock 181--190,
\newblock \urlprefix\url{https://doi.org/10.1145/3183519.3183525}.

\bibitem{Sako:2021}
\newblock M.~Sako,
\newblock From remote work to working from anywhere,
\newblock \emph{Communications of the {ACM}}, \textbf{64} (2021), 20--22,
\newblock
  \urlprefix\url{https://cacm.acm.org/magazines/2021/4/251338-from-remote-work-to-working-from-anywhere/fulltext}.

\bibitem{Scarnati:2001}
\newblock J.~T. Scarnati,
\newblock On becoming a team player,
\newblock \emph{Team Performance Management: An International Journal},
  \textbf{7} (2001), 5--10,
\newblock \urlprefix\url{https://doi.org/10.1108/13527590110389501}.

\bibitem{Sheffield:2013}
\newblock J.~Sheffield and J.~Lem{\'e}tayer,
\newblock Factors associated with the software development agility of
  successful projects,
\newblock \emph{International Journal of Project Management}, \textbf{31}
  (2013), 459--472,
\newblock \urlprefix\url{https://doi.org/10.1016/j.ijproman.2012.09.011}.

\bibitem{Stewart:2006}
\newblock G.~L. Stewart,
\newblock A meta-analytic review of relationships between team design features
  and team performance,
\newblock \emph{Journal of Management}, \textbf{32} (2006), 29--55,
\newblock \urlprefix\url{https://doi.org/10.1177/0149206305277792}.

\bibitem{The-Standish-Group:2018}
\newblock {The Standish Group},
\newblock Comprehensive {H}uman {A}ppraisal for {O}riginating {S}oftware
  ({CHAOS}) report 2018: Decision latency theory: It's all about the interval,
  2018,
\newblock \urlprefix\url{https://www.standishgroup.com}.

\bibitem{Trytten:2005}
\newblock D.~A. Trytten,
\newblock A design for team peer code review,
\newblock \emph{ACM SIGCSE Bulletin}, \textbf{37} (2005), 455--459,
\newblock \urlprefix\url{https://doi.org/10.1145/1047124.1047492}.

\bibitem{Valentin:2015}
\newblock E.~Valentin, J.~R. Hughes~Carvalho and R.~Barreto,
\newblock Rapid improvement of students' soft-skills based on an agile-process
  approach,
\newblock in \emph{{2015 IEEE Frontiers in Education Conference (FIE)}}, 2015,
\newblock 1--9,
\newblock \urlprefix\url{https://doi.org/10.1109/FIE.2015.7344408}.

\bibitem{Mierlo:2006}
\newblock H.~van Mierlo, C.~G. Rutte, J.~K. Vermunt, M.~A.~J. Kompier and J.~A.
  M.~C. Doorewaard,
\newblock Individual autonomy in work teams: The role of team autonomy,
  self-efficacy, and social support,
\newblock \emph{European Journal of Work and Organizational Psychology},
  \textbf{15} (2006), 281--299,
\newblock \urlprefix\url{https://doi.org/10.1080/13594320500412249}.

\bibitem{Vazquez-Bustelo:2007}
\newblock D.~V{\'a}zquez‐Bustelo, L.~Avella and E.~Fern{\'a}ndez,
\newblock Agility drivers, enablers and outcomes: Empirical test of an
  integrated {A}gile manufacturing model,
\newblock \emph{International Journal of Operations \& Production Management},
  \textbf{27} (2007), 1303--1332,
\newblock \urlprefix\url{https://doi.org/10.1108/01443570710835633}.

\bibitem{Yost:2000}
\newblock C.~A. Yost and M.~L. Tucker,
\newblock Are effective teams more emotionally intelligent? confirming the
  importance of effective communication in teams,
\newblock \emph{Delta Pi Epsilon Journal}, \textbf{42} (2000), 101--109.

\bibitem{Zahedia:2016}
\newblock M.~Zahedia, M.~Shahinb and M.~A. Babar,
\newblock A systematic review of knowledge sharing challenges and practices in
  global software development,
\newblock \emph{International Journal of Information Management}, \textbf{36}
  (2016), 995--1019,
\newblock \urlprefix\url{https://doi.org/10.1016/j.ijinfomgt.2016.06.007}.

\bibitem{Zhang:2020}
\newblock A.~X. Zhang, M.~Muller and D.~Wang,
\newblock How do data science workers collaborate? roles, workflows, and tools,
\newblock \emph{Proceedings of the {ACM} on Human-Computer Interaction},
  \textbf{4} (2020), Article 22,
\newblock \urlprefix\url{https://doi.org/10.1145/3392826w}.

\bibitem{Zhang:2010}
\newblock S.~Zhang, F.~K{\"o}bler, M.~Tremaine and A.~Milewski,
\newblock Instant messaging in global software teams,
\newblock \emph{International Journal of e-Collaboration}, \textbf{6} (2010),
  43--63,
\newblock \urlprefix\url{https://doi.org/10.4018/978-1-61350-459-8.ch010}.

\bibitem{Zolkifli:2018}
\newblock N.~N. Zolkifli, A.~Ngah and A.~Deraman,
\newblock Version control system: A review,
\newblock \emph{Procedia Computer Science}, \textbf{135} (2018), 408--415,
\newblock \urlprefix\url{https://doi.org/10.1016/j.procs.2018.08.191}.

\end{thebibliography}

\medskip
Received xxxx 20xx; revised xxxx 20xx.
\medskip

\end{document}